\documentclass[a4paper,11pt]{article}

\usepackage{amsfonts,amssymb}
\usepackage{theorem}
\usepackage{epsf}
\usepackage{eufrak}

\def\G{\Gamma}

\newtheorem{theorem}{Theorem}[section]
\newtheorem{prop}{Proposition}[section]
\newtheorem{lemma}{Lemma}[section]
\newtheorem{definition}{Definition}[section]
\newtheorem{definizione}{Definizione}[section]
\newtheorem{corollary}{Corollary}[section]
\newtheorem{note}{Note}[section]
\begin{document}

\rightline{IFUM-800-FT and MPP-2004-86}

\vskip 1.3 truecm
\Large
\bf
\centerline{Physical Unitarity for Massive Non-abelian Gauge}
 \par \centerline{ Theories in the Landau Gauge: St\"uckelberg 
\& Higgs } 
\normalsize \rm

\vskip 0.5 truecm
\large
\centerline{Ruggero Ferrari $^a$
\footnote{E-mail address: {\tt ruggero.ferrari@mi.infn.it}} 
and Andrea Quadri $^b$ \footnote{E-mail address: {\tt quadri@mppmu.mpg.de}}}

\vskip 0.3 truecm
\normalsize
\centerline{$^a$Phys. Dept. University of Milan, 
via Celoria 16, 20133 Milan, Italy } 
\centerline{I.N.F.N., sezione di Milano} 
\centerline{$^b$Max-Planck-Institut f\"ur Physik (Werner-Heisenberg-Institut)} 
\centerline{F\"ohringer Ring, 6 - D80805 M\"unchen, Germany}

\vskip 0.5  truecm
\normalsize
\bf
\centerline{Abstract}
\rm
\begin{quotation}
We discuss the problem of unitarity for 
Yang-Mills theory in the Landau gauge with a mass term 
{\sl \`a la} St\"uckelberg. 
We assume that the theory (non-renormalizable) 
makes sense in some subtraction scheme (in particular the Slavnov-Taylor 
identities should be respected!) and we devote the paper to the study of 
the space of the unphysical modes. We find that the theory is unitary only 
under the hypothesis that the
1-PI two-point function of the vector mesons
has no poles (at $p^2=0$). This normalization condition might be rather
crucial in the very definition of the theory.  
With all these provisos the theory is unitary. The proof of unitarity 
is given both in a form that allows a direct transcription in terms of Feynman 
amplitudes (cutting rules) and in the operatorial form.
\par
The same arguments and conclusions apply {\sl verbatim} to the case
of non-abelian gauge theories where the mass of the vector meson
is generated via Higgs mechanism. To the best of our knowledge, there is
no mention in the literature on the necessary condition implied
by physical unitarity.

\end{quotation}

\newpage

\vskip 0.8 truecm

\section{Introduction}
\label{sec:intr}

The quest for a consistent non-abelian gauge theory~\cite{ym} of massive
gauge bosons is a subject with a long and venerable 
history. Today the preferred
solution, combining unitarity and renormalizability, is still the 
spontaneous symmetry breaking mechanism based on the introduction
of the Higgs field \cite{Higgs:1964ia}.
Nevertheless, within the context of non-power-counting renormalizable
models, the St\"uckelberg mechanism 
\cite{stueckelberg} has been repeatedly advocated~\cite{ruegg,Esole:2004rx}
as a possible alternative for the generation of massive
 non-abelian vector fields.

This paper is devoted to the discussion of 
some crucial points in the proof of Physical Unitarity for a massive 
non-abelian gauge theory 
in the presence of a mass term {\sl \`a la }St\"uckelberg.
Such term has been originally introduced in order to have a gauge invariant
theory for  massive photons. It can be seen as the result of an operatorial
gauge transformation on the fields in the Proca gauge. The same procedure
can be envisaged also in the case of non-abelian gauge theories
\cite{Slavnov:1970tk}-
\cite{delbourgo}.
While in the abelian case the theory is renormalizable and moreover
the proof of unitarity on the physical states poses no problems, 
the non-abelian
case is far more complicated. The origin of the troubles is mainly
the term generated in the mass by the
operatorial
gauge transformation:
it yields a non-polynomial lagrangian. The efforts made in order to overcome
these difficulties is a long list in the history of quantum field theory.
Along this line one of the first steps to be accomplished
is a close analysis of the
problem of unitarity. In fact even if the theory is made finite by
some subtraction scheme, physical unitarity will always be one crucial
item to be considered. The present work is aimed at focusing on the
conditions that have to be met in order to guarantee this important
property. Previous attempts to prove unitarity 
made use of a
direct diagrammatic study of the Feynman amplitudes and they are limited
to one-loop in the perturbative expansion \cite{delbourgo}.

\medskip
The classical lagrangian for $SU(2)$ is
\begin{eqnarray}&&
{\cal L}= -\frac{1}{4}G_{a\mu\nu}G_a^{\mu\nu}
+m^2 ~Tr~\left(
A_\mu + \frac{i}{g}\Omega\partial_\mu\Omega^\dagger
\right)^2+ {\cal L}_M ,
\nonumber\\&&
G_{a\mu\nu}=\partial_\mu A_{a\nu}-\partial_\nu A_{a\mu}+\epsilon_{abc}A_{b\mu}
A_{c\nu},
\label{int.0.1}
\end{eqnarray}
where $m$ is the St\"uckelberg mass 
and $\Omega$ is parameterized  in terms of the
St\"uckelberg fields $\vec{\phi}$ by
\begin{eqnarray}
m \Omega = \phi_0 {\bf 1} + i \vec{\phi} \cdot \vec{\tau} \, 
\label{int.0.1.bis}
\end{eqnarray}
with $\phi_0 = \sqrt{m^2 - \phi_a^2}$.

The discussion is devoted  to the particular formulation provided 
by the Landau gauge-fixing term 
\begin{eqnarray}
S_{g.f.} = \int d^4 x ~2\,Tr\left(
B\partial^\mu A_\mu - \bar c \partial^\mu D[A]_\mu c
\right).
\label{int.1}
\end{eqnarray}
We chose this gauge because with a transverse vector field
propagator the Feynman rules are particularly simple. In particular
there are no important out-of-diagonal terms in the connected
two-point functions. 

In Appendix \ref{app:other} we elaborate
upon other covariant gauges and demonstrate that the subspace of
the unphysical modes includes also dipole fields, as it is known
in power-counting renormalizable theories.

\par
The presence of a further scalar mode, introduced via the 
St\"uckelberg mass term, requires a revisitation of the standard
proof of physical unitarity in non-abelian gauge theories \cite{Becchi:1975nq},
\cite{Becchi:1974xu}, \cite{Curci:1976yb}, \cite{Kugo:1977zq}.
In particular a detailed study of the Fock space is necessary
in order to identify the unphysical modes. The usual method
based on the study of the kernel of the BRST charge 
\cite{Curci:1976yb} \cite{Kugo:1977zq}
\begin{eqnarray} 
|Phys\rangle  \in {\rm Ker } ~ Q / {\rm Im} ~ Q \,
\label{int.0.2} 
\end{eqnarray}
has to be supported by a preliminary study of the Fock
space of the theory. In particular it appears that too many fields
describe asymptotically massless unphysical modes (the vector field, the
Nakanishi-Lautrup field ~\cite{Nakanishi:sm}\cite{lautrup}
 and the St\"uckelberg field). A condition
has to be met in order that the definition of physical space in eq.
(\ref{int.0.2}) guarantees physical unitarity.
The Landau gauge requires that the connected two-point
function for the gauge bosons is transverse
\begin{eqnarray}
W_{ab}^{\mu\nu} = \delta_{ab} W_T(p^2)\left(g^{\mu\nu}-
\frac{p^\mu p^\nu}{p^2} \right). 
\label{int.2} 
\end{eqnarray} 
The main result of the paper is the following. 
If one can make sense out of a non-abelian gauge theory with 
a St\"uckelberg mass term, then the physical unitarity is 
satisfied provided one can impose the normalization condition 

\begin{eqnarray} 
W_T (0)= \lim_{p^2=0} \frac{W_{\phi \phi}}{p^2 W_{B \phi}^2}, 
\label{int.3} 
\end{eqnarray} 
where $\phi$ is the St\"uckelberg field.
The importance of the result is quite transparent: if the 
condition cannot be enforced, unitarity (for the physical states) 
is lost. Thus this seems to be the very crucial condition to meet 
in order to define the theory.
\par
The same discussion and the same condition (\ref{int.3}) is valid for a massive
non-abelian gauge theory where the mass is generated by the Higgs
mechanism. In this case the r\^ole of the St\"uckelberg field
is played by the unphysical components of the Higgs field. 
The proof here provided for the
St\"uckelberg case is valid also for the Higgs case since the
ingredients used are essentially the same: i) the Slavnov-Taylor identities
~\cite{st},
ii) the equation of motion of the Nakanishi-Lautrup field
and iii) the equation of motion of the ghost field~\cite{fp}. After this
remark we have chosen to discuss only the St\"uckelberg
case since we optimistically hope that some day the obstacle of
non-renormalizability~\cite{Boulware} will be removed and a  theory, consistent
from the point view of physics,
will be thus achieved.

\medskip
The paper is organized as follows. The abelian St\"uckelberg
model is reviewed in Sect.~\ref{sec:abel}.
The non-abelian case is taken up in Sect.~\ref{sec:nonabel}.
The diagrammatic approach
\cite{'tHooft:1971rn} to the analysis of physical unitarity
is presented in Sect.~\ref{sec:diagr}, while the complementary operatorial
approach \cite{Curci:1976yb,Kugo:1977zq,Becchi:1985bd} 
is considered in Sect.~\ref{sec:operatorial}. Finally conclusions
are given in Sect.~\ref{sec:conclusions}.


\section{Abelian case}
\label{sec:abel}
As a warm-up we consider the Landau gauge in the abelian case.
The Lagrangian is
\begin{eqnarray} {\cal L} = -\frac{1}{4}F_{\mu\nu}^2 + \frac{m^2}{2}(A_\mu
-\frac{1}{m} \partial_\mu\phi)^2
+B\partial_\mu A^\mu  + {\cal L}_M \, , 
\label{abel.1}
\end{eqnarray}
where matter enters in ${\cal L}_M$.
For abelian theories the St\"uckelberg mechanism leads to a power-counting
renormalizable model.

The Ward identity is (dots indicate the part relevant for matter) 
\begin{eqnarray} \Box  W_{B} +\partial_\mu J_A^\mu
- mJ_\phi+\dots =0
\label{abel.2}
\end{eqnarray}
the B equation of motion\footnote{The notation is as follows: $W_\psi$ stands
for $\delta W / \delta J_\psi$, with $\psi$ any of the quantized
fields of the model and $J_\psi$ its source. The connected
generating functional $W[J_\psi]$ is related to the vertex functional
$\G[\psi]$ by $W=\G+ \int d^4x \, J_\psi \psi$ \, .}
\begin{eqnarray} 
J_B+\partial^\mu W_{A^\mu} =0 \, , 
\label{abel.3}
\end{eqnarray}
and the $\phi$ equation of motion
\begin{eqnarray}
J_\phi+ m(\partial^\mu  W_{A^\mu}
- \frac{1}{m} \Box W_{ \phi}  )=0.
\label{abel.3.1}
\end{eqnarray}
It is useful to have these equations for the vertex functional
\begin{eqnarray} &&
\Box B -\partial_\mu\Gamma_{A_\mu} +m \Gamma_{\phi}+\dots=0 \label{abel.3.3a} 
\\
&& \Gamma_B = \partial^\mu A_\mu \label{abel.3.3b} \\
&& \Gamma_{\phi}= m(\partial_\mu A^\mu-\frac{1}{m}\Box \phi ). 
\label{abel.3.3c} 
\end{eqnarray} 
One obtains easily 
\begin{eqnarray}&& 
W_{A^\mu \phi}= 0, \quad
W_{\phi \phi}=- \frac{1}{p^2} , \quad
W_{\phi B}= \frac{m}{ p^2} , \quad
\nonumber\\&&
W_{B B}= 0 , \quad
W_{A^\mu(p) B}= - i \frac{p_\mu}{ p^2}.
\label{abel.4}
\end{eqnarray} 
The propagator of the gauge bosons is transverse
\begin{eqnarray}
W_{A_\mu A_\nu} = \frac{1}{p^2 - m^2} T_{\mu \nu} \, .
\label{abel.prop}
\end{eqnarray}
where $T_{\mu\nu}$ is the projector
\begin{eqnarray}
T_{\mu\nu} = g_{\mu\nu} - \frac{p_\mu p_\nu}{p^2} \, .
\label{abel.proj}
\end{eqnarray}
We will also need later on the orthogonal longitudinal projector
$L_{\mu\nu} = \frac{p_\mu p_\nu}{p^2}$. 
The final goal of the calculation is the construction of three 
linearly independent
fields,
spanning the bosonic sector at $p^2=0$, 
such that at the pole in $p^2 = 0 $ one has
\begin{eqnarray}&&  
W_{BB}= 0, \quad
W_{BX}= \frac{1}{ p^2} , \quad
W_{XX}= 0 , \quad
\nonumber\\&& 
W_{Y^\mu A^\nu}= 0 , \quad 
W_{Y^\mu B}= 0 , \quad
W_{Y^\mu X}= 0. 
\label{abel.5} 
\end{eqnarray} 
By using eqs. (\ref{abel.4}) one gets 
\begin{eqnarray} X =  \frac{1}{2m^2} B + \frac{1}{m}
\phi. \label{abel.6} \end{eqnarray} 
Now we construct the field $Y^\mu$ 
\begin{eqnarray} Y^\mu = a A^\mu + b \partial^\mu  B + c \partial^\mu X 
\label{abel.7} 
\end{eqnarray} 
and we get
\begin{eqnarray}&&
W_{Y^\mu A^\nu}= a W_{A^\mu A^\nu} + b \frac{p_\mu p_\nu}{p^2} 
+ c \frac{p_\mu p_\nu}{2m^2p^2}=0
\label{abel.8a}\\&&
 W_{Y^\mu B}= i a\frac{p_\mu }{p^2} +i c \frac{p_\mu }{p^2} =0 
\label{abel.8b}\\&&  W_{Y^\mu X}=  i a\frac{p_\mu }{2m^2p^2} +ib 
\frac{p_\mu }{p^2}=0 \, ,
\label{abel.8c} \end{eqnarray} 
 which admits a non-trivial solution only if the pole part of $W_{A_\mu A_\nu}$
at $p^2=0$ is
\begin{eqnarray} W_{A^\mu A^\nu} = \frac{p_\mu p_\nu}{m^2p^2} \qquad 
{\rm for}\quad p^2 \sim 0 
\label{abel.9} 
\end{eqnarray} 
and then 
\begin{eqnarray}&& 
Y^\mu = a\left( A^\mu -  \frac{1}{2m^2}\partial^\mu  B - \partial^\mu X
\right)
\nonumber\\&&
= a\left( A^\mu -  \frac{1}{m^2}\partial^\mu  B -
\frac{1}{m} \partial^\mu \phi \right). 
\label{abel.10} 
\end{eqnarray} 
\par 
The condition in eq. (\ref{abel.9}) is related to a further property of 
the theory. Consider the vertex functional $\Gamma$. 
For the vector field we have 
\begin{eqnarray} \Gamma_{\mu\nu} = 
\Gamma_T \Big (g_{\mu\nu}- \frac{p_\mu p_\nu}{p^2} \Big )
+ \Gamma_L \frac{p_\mu p_\nu}{p^2}.
\label{abel.11}
\end{eqnarray}
From the definition the two-point functions obey the equation
\begin{eqnarray}
W \Gamma= - 1.
\label{abel.12}
\end{eqnarray}
Then we get
\begin{eqnarray}
W_T\Gamma_T= -1.
\label{abel.12.1}
\end{eqnarray}
and
\begin{eqnarray}
W_{B A^\mu} \Gamma_{A_\mu A_\nu}+W_{B\phi} \Gamma_{\phi A_\nu}= 0. 
\label{abel.13} 
\end{eqnarray} 
By using eqs.(\ref{abel.3.3c}) and (\ref{abel.4}) 
\begin{eqnarray} 
i\frac{p_\mu}{p^2}\Gamma_{A_\mu A_\nu}
+ \frac{1}{p^2}(-i m^2 p^\nu)= 0 
\label{abel.14} 
\end{eqnarray} 
i.e. 
\begin{eqnarray} \Gamma_L =  m^2 \label{abel.15} \end{eqnarray} %
and finally %
\begin{eqnarray} \Gamma_{\mu\nu} = -\frac{1}{W_T}
\Big (g_{\mu\nu}- \frac{p_\mu p_\nu}{p^2} \Big )
+ m^2 \frac{p_\mu p_\nu}{p^2}.
\label{abel.16}
\end{eqnarray}
The absence of singularities in $\Gamma$ at $p^2 = 0$ requires
\begin{eqnarray}
W_T(0)= - \frac{1}{m^2}
\label{abel.17}
\end{eqnarray}
i.e. the condition in eq. (\ref{abel.9}).
This condition becomes a non-trivial normalization condition
if one introduces matter in the theory. The radiative corrections
to the vector field propagator can be described by the 1-PI vertex
function
\begin{eqnarray}
\Pi^{\mu\nu}(p)=\Pi(p^2, M)(p^2 g^{\mu\nu}- p^\mu p^\nu)
\label{abel.18}
\end{eqnarray}
Condition (\ref{abel.17}) requires
\begin{eqnarray}
\lim _{p^2=0} p^2\Pi(p^2, M) =0,
\label{abel.19}
\end{eqnarray}
which means a mild behavior of $\Pi$ in zero. 

\section{The non-abelian case}
\label{sec:nonabel}
We consider the internal group $SU(2)$ for sake of simplicity. 
We choose the generators of the Lie algebra $su(2)$ to be
$T^a = \frac{1}{2} \tau^a$, with $\tau^a$ the Pauli matrices.
Then 
$[T^a,T^b] = i \epsilon^{abc} T^c \, .$
In the non-abelian case the Ward identity is replaced by the Slavnov-Taylor 
identity and the r\^ole of the equation of motion of the St\"uckelberg 
field is taken up by the equation of motion of the ghost field.
At tree-level the action of Yang-Mills theory with a St\"uckelberg mass
is
\begin{eqnarray}
S & = & \int d^4x \, \Big (
-\frac{1}{4} G_{\mu\nu}^a G^{a\mu\nu } +
\frac{m^2}{2} \, \Big ( A^a_\mu + 
F^a_\mu \Big )^2 
\nonumber \\
& & ~~~~~~~~~+ B^a \partial A^a - \bar c^a \partial_\mu D^\mu[A] c^a 
\Big ) \, ,
\label{nonabel.gen.3}
\end{eqnarray}
where the field strength $G_{\mu\nu}^a$ is given by
\begin{eqnarray}
G_{\mu\nu}^a = \partial_\mu A_\nu^a - \partial_\nu A_\mu^a 
+ \epsilon^{abc} A_\mu^b A_\nu^c \, 
\label{nonabel.gen.4}
\end{eqnarray}
and the pure gauge vector field $F_\mu$ is
\begin{eqnarray}
F_\mu = F_\mu^a T^a = \frac{i}{2} \Omega \partial_\mu \Omega^\dagger \, .
\label{nonabel.gen.5}
\end{eqnarray}
$\Omega$ is a unitary matrix parameterized by the fields $\phi^a$:
\begin{eqnarray}
m \Omega = \phi_0 {\bf 1} + i \vec{\phi} \cdot \vec{\tau}
\label{nonabel.gen.6}
\end{eqnarray}
with
$\phi_0=\sqrt{m^2-\phi_a^2}$. 
%

The Slavnov-Taylor identities are derived from the BRST
\cite{BRST} 
invariance of the action 
\begin{equation}
\begin{array}{lll}
s~ A_{a\mu} = D[A]_{ab\mu} c_b, & s~ \bar c_a = B_a, &
s~\phi_a = \frac{1}{2}\left (c_a \phi_0  -\epsilon_{abc}c_b \phi_c\right),\\
s~ c_a = -\frac{1}{2}\epsilon_{abc}c_b c_c, & s~B_a = 0, &
s~\phi_0 = -  \frac{1}{2} c_a\phi_a 
\end{array}
\label{nonabel.1}
\end{equation}
i.e. (matter will be omitted,
group indices and integration over space-time are understood)
\begin{eqnarray}
-~{\cal S}(W) &=& J_{A}^\mu W_{A^{*\mu}} + J_{\phi}W_{\phi^*}+ J_{\bar c}W_{B}
+J_{c}W_{c^*}=0
\label{nonabel.2a}
\\
{\cal S}(\G) &=& \Gamma_{A_\mu}\Gamma_{A^{*\mu}} + \Gamma_{\phi}\Gamma_{\phi^*}
+ B \Gamma_{\bar c}
+\Gamma_{c}\Gamma_{c^*}=0.
\label{nonabel.2b}
\end{eqnarray}

A star superscript over a field variable denotes the corresponding
antifield \cite{zj, Gomis:1994he}.
The relevant local cohomology of the linearized classical
Slavnov-Taylor operator has been studied in \cite{Henneaux:1998hq}.

The $B-$field equation of motion is
\begin{eqnarray}&&
J_{B}+\partial^\mu W_{A^{\mu}}=0
\label{nonabel.3a}
\\&&
\Gamma_{B}= \partial^\mu A_{\mu}.
\label{nonabel.3b}
\end{eqnarray}
The ghost equation of motion is
\begin{eqnarray}&&
- J_{\bar c} + \partial^\mu W_{A^{*\mu}}=0
\label{nonabel.4a}
\\&&
\Gamma_{\bar c}= -\partial^\mu\Gamma_{ A^{*\mu}}. \label{nonabel.4b} \end{eqnarray} 
From the eqs. (\ref{nonabel.2a}-\ref{nonabel.4b}) one gets 
\begin{eqnarray}
&& W_{A^\mu B} = -i\frac{p_\mu}{p^2}, \quad
W_{A^\mu \phi} =0, \quad
W_{B B} = 0, \quad
\nonumber \\&& 
W_{B \phi} = W_{\phi^* \bar c}, \quad 
W_{A^{*\mu}\bar c}= i\frac{p_\mu}{p^2} 
\label{nonabel.5} 
\end{eqnarray} 
and similarly 
\begin{eqnarray}&& 
\Gamma_{B A^\mu } = -i p_\mu, \quad 
\Gamma_{B \phi} =0 , \quad 
\Gamma_{B B} = 0  , \quad 
\nonumber \\&& 
\Gamma_{\bar c c} =  -i p^\mu \Gamma_{A^{*\mu}c} , \quad 
\Gamma_{A_\mu\phi}\Gamma_{A^{*\mu}c} + \Gamma_{\phi\phi}\Gamma_{\phi^*c} = 0.
\label{nonabel.6}
\end{eqnarray}

Eq. (\ref{nonabel.6}) implies
\begin{eqnarray}
W_{A^{*\mu}\bar c} = \Gamma_{A^{*\mu}c} W_{c \bar c} 
= \Gamma_{A^{*\mu}c} \Gamma_{c \bar c}^{-1} = i\frac{p_\mu}{p^2} . 
\label{nonabel.7} 
\end{eqnarray} 

Now we use the above equations to get the longitudinal part
of the two-point vertex function. 
We consider the relevant components of the matrix product 
\begin{eqnarray} 
W \Gamma= - 1. 
\label{nonabel.8} 
\end{eqnarray}
We get 
\begin{eqnarray}&& 
W_{A^\mu A^\rho}\Gamma_{A_\rho A^\nu } 
+W_{A^\mu B}\Gamma_{B A^\nu }
= -g_{\mu\nu}\quad \Longrightarrow W_T \Gamma_T = -1 
\label{nonabel.9a} \\&& 
W_{A^\mu A^\rho}\Gamma_{A_\rho B} 
+W_{A^\mu \phi}\Gamma_{\phi B }=0 
\label{nonabel.9b} \\&& 
W_{A^\mu A^\rho}\Gamma_{A_\rho \phi} 
+W_{A^\mu \phi}\Gamma_{\phi \phi }=0 
\label{nonabel.9c} \\&& 
W_{B A^\rho}\Gamma_{A_\rho A^\mu } +W_{ B\phi}\Gamma_{\phi A^\mu }=0 
\quad\Longrightarrow
%
 i \Gamma_L \frac{p_\mu}{p^2} 
+  W_{ B\phi}\Gamma_{\phi A^\mu }=0 
\label{nonabel.9d} \\&& 
W_{B A^\rho}\Gamma_{A_\rho B } +W_{ B\phi}\Gamma_{\phi B }= -1 
\label{nonabel.9e} \\&& 
W_{ BA_\mu}\Gamma_{A^\mu\phi} + W_{B \phi}\Gamma_{\phi \phi}=0 
\label{nonabel.9f} \\&& 
W_{\phi B}\Gamma_{BA^\mu} + W_{\phi \phi}\Gamma_{\phi A^\mu}=0 \quad 
\Longrightarrow
-ip_\mu W_{ B\phi} + W_{\phi \phi}\Gamma_{\phi A^\mu}=0 
\label{nonabel.9g} \\&& 
W_{\phi A^\rho}\Gamma_{A_\rho B } +W_{\phi B}\Gamma_{B B}
+W_{\phi \phi}\Gamma_{\phi B}=0
\label{nonabel.9h}
\\&&
W_{\phi A^\rho}\Gamma_{A_\rho \phi } +W_{\phi B}\Gamma_{B \phi}
+W_{\phi \phi}\Gamma_{\phi \phi}=-1 \quad  \Longrightarrow
W_{\phi \phi}\Gamma_{\phi \phi}=-1
\label{nonabel.9i}
\end{eqnarray}
From eqs. (\ref{nonabel.9d}) and (\ref{nonabel.9g}) we get
\begin{eqnarray}
\Gamma_L = -\frac{p^2 W_{B \phi}^2 }{W_{\phi \phi}}. 
\label{nonabel.10} 
\end{eqnarray} 
The requirement that $\Gamma$ has no poles in $p^2 = 0$ gives 

\begin{eqnarray} 
\lim_{p^2=0}(\Gamma_L - \Gamma_T)=0. 
\label{nonabel.11} 
\end{eqnarray} 
Then from eqs.  (\ref{nonabel.9a}), (\ref{nonabel.10}) and 
(\ref{nonabel.11}) 
\begin{eqnarray} 
\lim_{p^2=0}\left(W_T -  \frac{W_{\phi \phi}}{p^2 W_{B \phi}^2 } \right)=0.
\label{nonabel.12} 
\end{eqnarray} 

\subsection{Construction of the unphysical modes}

As in the previous section we construct the fields that describe 
conveniently the unphysical modes 
\begin{eqnarray} 
X 
=-\left .\frac{W_{\phi \phi}}{2p^2W_{B \phi}^2}\right|_{p^2=0}  B
+\left . \frac{1}{p^2W_{B \phi}}\right|_{p^2=0} \phi
= - \frac{1}{2}W_T(0)  B
+\left . \frac{1}{p^2W_{B \phi}}\right|_{p^2=0} \phi.
\label{nonabel.13}
\end{eqnarray}
The existence of a linear combination of fields that has no pole at $p^2=0$

\begin{eqnarray}
Y^\mu = a A^\mu + b \partial^\mu  B + c \partial^\mu X 
\label{nonabel.14} 
\end{eqnarray} 
requires
\begin{eqnarray}&&
W_{Y^\mu A^\nu}= a W_{A^\mu A^\nu} + b \frac{p_\mu p_\nu}{p^2} 
- c \frac{p_\mu p_\nu}{p^2}\frac{W_T(0)}{2}=0
\nonumber\\&&
 W_{Y^\mu B}=  -i a\frac{p_\mu }{p^2} -i c \frac{p_\mu }{p^2} =0 %
\nonumber\\&&
 W_{Y^\mu X}=   i a\frac{p_\mu }{p^2}\frac{W_T(0)}{2}
 -ib \frac{p_\mu }{p^2}  =0.
\label{nonabel.15}
\end{eqnarray}
This set of equations has a non-trivial solution only if 
\begin{eqnarray} 
W_T (0)= \lim_{p^2=0} \frac{W_{\phi \phi}}{p^2 W_{B \phi}^2} 
\label{nonabel.16} 
\end{eqnarray} 
which is guaranteed by the requirement in eq. (\ref{nonabel.12})
and it is the condition necessary in order that the determinant
of the residuum
\begin{eqnarray} {\cal R}_{ij}(p) = \lim_{p^2=0} p^2W_{ij}(p) 
\label{nonabel.16.1} 
\end{eqnarray} 
is zero.
Thus 
\begin{eqnarray}&& Y^\mu = a \left(A^\mu 
+ \frac{W_T(0)}{2 }\partial^\mu  B - \partial^\mu X
\right)
\nonumber\\&&
=a \left(A^\mu  +W_T(0) \partial^\mu  B
-  \left . \frac{1}{p^2W_{B \phi}}\right|_{p^2=0}
\partial^\mu \phi
\right).
\label{nonabel.17}
\end{eqnarray}
Thus we can choose another doublet of fields. 
Instead of $B,X$ one can use the scalar part of the fields 
\begin{eqnarray}&& B^\mu \equiv 
\frac{1}{W_T(0)}\left(-A^\mu 
+ \left . \frac{1}{p^2W_{B \phi}}\right|_{p^2=0} \partial^\mu \phi
\right)
\label{nonabel.18a}\\&&
X^\mu\equiv 
\frac{1}{2}\left(A^\mu + \left . \frac{1}{p^2W_{B \phi}}\right|_{p^2=0} 
\partial^\mu \phi
\right) .
\label{nonabel.18b}
\end{eqnarray}


%
\section{Diagrammatic approach}
\label{sec:diagr}
The results of the previous sections allow us to
trace the identities for a diagrammatic approach to the
problem of physical unitarity \cite{Cutkosky:1960sp}\cite{Veltman:1963th} . 
The unphysical states are expected to be massless modes. 
Few elements support this statement. 
In particular the $B-$ equation of motion in (\ref{nonabel.3a}) 
states that the connected two-point function of the vector field 
is pure transverse and therefore it contains a pole at zero mass 
describing a scalar particle. Consequently if we assume that we have 
asymptotic states the BRST transformations in eq. (\ref{nonabel.1}) 
require that the fields $B,\phi$ describe massless modes. 
We have seen in the previous section that only two boson massless modes exist, 
with opposite metric. 
As a last comment before entering deeply in the calculation we notice 
that due to eq. (\ref{nonabel.5}) the $B-$field never appear 
as an internal line in the actual calculation of Feynman amplitudes. 
In verifying physical unitarity directly on amplitudes 
we have to consider the projection operators corresponding to the massless 
modes described by the fields $A_\mu, \phi, c, \bar c$. 
Due to the eq. (\ref{nonabel.5}), there is no mixing between 
the $A_\mu$ and the $\phi$ fields. 
\par
Let us formulate the problem in a schematic way. We have
a set of fields
\begin{eqnarray}
\psi_i = \{A_{a\mu},\phi_a,B_a,c_a,\bar c_a,\dots \} 
\label{diagr.1} 
\end{eqnarray} 
The two-point function can be expanded around the poles 
\begin{eqnarray}
W_{ij}(x) = \sum_\lambda \int d^3 p 
\frac{1}{2E_\lambda} \langle 0 |\psi_i(x)|\lambda \vec p\rangle 
\rho^{-1}_\lambda \langle\lambda \vec p|\psi_j(0)|0\rangle,
\label{diagr.2} 
\end{eqnarray} 
where the index $\lambda$ collects all internal indices 
(the mass is $m_\lambda$).
The states are normalized by 
\begin{eqnarray} \langle\lambda \vec p\,|\lambda'\vec  p\,'\rangle 
= 2 E_p \rho_\lambda\delta_{\lambda \lambda' } \delta_3(p-p') 
\label{diagr.3} 
\end{eqnarray} 
where $\rho_\lambda$ is the metric tensor. 
The {\sl wave functions} are introduced 
\begin{eqnarray} f_{\lambda p i}(x) = 
\langle 0 |\psi_i(x)|\lambda \vec p\rangle 
\label{diagr.4} 
\end{eqnarray} 
and on them it is convenient to introduce the bilinear form 
(all asymptotic fields obey Klein-Gordon equations) 
\begin{eqnarray} \left(g_{\lambda '}, f_{\lambda }\right) = i\rho_\lambda
\delta_{\lambda \lambda' }  \sum_i \int d^3 x g_{\lambda 'i}^* 
\stackrel{\leftrightarrow}{\partial}_0 f_{\lambda i}. 
\label{diagr.5} 
\end{eqnarray} 
The wave functions form a vector space $\cal V$. 
The form in eq. (\ref{diagr.5}) allows to define a dual space 
$\widetilde{\cal V}$. 
In particular we can define an {\sl orthogonal} function 
$\tilde f_{\lambda p} \in \widetilde{\cal V}$ for any
$f_{\lambda p} \in {\cal V}$ such that
\begin{eqnarray}
\left(\tilde f_{\lambda' p'}, f_{\lambda p}\right) 
= \rho_\lambda \delta_{\lambda \lambda' } 2 E_p \delta_3(p-p'). 
\label{diagr.6} 
\end{eqnarray} 
This unusual structure is necessary in order to deal with a degenerate 
scalar product in $\cal V$. For instance there are wave functions 
of 
gradients of a massless scalar field. 
Moreover there are more fields than massless modes. 
\par
Now we are in place to define the S-matrix  elements in terms of the 
connected amplitude by using the operator (for an outgoing mode)
\begin{eqnarray}&&
\langle\dots, \vec p \lambda,\dots|S|\dots\rangle = 
\nonumber\\&& \dots\left( -\sum_{ij} \int d^4 x f_{\lambda pi}^*(x)d^4 y 
\Gamma_{ij}(x-y) \frac{\delta}{\delta J_j(y)}\right)
\dots \left .W[J]\right|_{J=0}. 
\label{diagr.7} 
\end{eqnarray} 
The above equation hints to introduce the operation of {\sl truncation} 
on the functional $W$ by 
\begin{eqnarray} W_{\widehat{\psi_i(p)}}[J] = 
- \lim_{p^2=m^2_\lambda}\int d^4 x d^4 y \exp(ipx)
\Gamma_{ij}(x-y)
\frac{\delta}{\delta J_j(y)} W[J]
\label{diagr.7.1}
\end{eqnarray}
i.e. external leg, starting with index $i$, is removed and the momentum 
is taken on-shell. The S-matrix element is recovered by 
folding in 
the wave function of the mode 
\begin{eqnarray} 
\langle\dots, \vec p \lambda,\dots|S|\dots\rangle =\dots \sum_i 
f_{\lambda pi}^*\left .W_{\widehat{\psi_i(p)}}[J]\right |_{J=0}. 
\label{diagr.7.2} 
\end{eqnarray} 
\par The more familiar reduction formula (LSZ-formalism) takes the form 
\begin{eqnarray} 
-i\sum_{k} \int d^4 x \tilde f_{\lambda pk}^*(x)\rho_\lambda 
(\Box + m^2_\lambda) \frac{\delta}{\delta J_k(x)}. 
\label{diagr.8} 
\end{eqnarray} 

\subsection{Diagrammatic unitarity}

Consider now the functional
\begin{eqnarray}
-\int d^4 x \exp(ipx)
\Box 
\frac{\delta}{\delta J_B(x)} \left . W[J]\right|_{p^2=0}
= \lim_{p^2=0} p^2W_{Bj}(p) W_{\widehat{\psi_j(p)}}[J] 
\label{diagr.9} 
\end{eqnarray} 
The relevant quantity is then the residuum 
\begin{eqnarray} {\cal R}_{ij}(p) = \lim_{p^2=0} p^2W_{ij}(p) 
\label{diagr.10} 
\end{eqnarray} 
and in particular eq. (\ref{diagr.9}) contains only %
\begin{eqnarray}&& {\cal R}_{BA^\mu}(p) = i p_\mu \nonumber\\&& 
{\cal R}_{B\phi}(p) = \lim_{p^2=0} p^2W_{B\phi^*}(p) 
\label{diagr.11} 
\end{eqnarray} 
as one sees from eq. (\ref{nonabel.5}). 
\par 
According to eq. (\ref{diagr.7}) 
in order to check the required cancellation among unphysical modes
in
the unitarity equation %
\begin{eqnarray} \langle f|i\rangle = 
\sum_n \langle f|S^\dagger|n\rangle \langle n|S|i\rangle 
\label{diagr.12} 
\end{eqnarray} 
we have to evaluate terms as (the dependence of $W$ on the external
currents is understood)
\begin{eqnarray}&&
\int \frac{d^3p}{2p}\left(W_{\widehat{A_\mu(p)}}^*
 {\cal R}_{A^\mu A^\nu}(p)W_{\widehat{A_\nu(p)}}  
+ 
W_{\widehat{\phi(p)}}^*
 {\cal R}_{\phi\phi}(p)W_{\widehat{\phi(p)}} 
\right)
\nonumber\\&&
=\int \frac{d^3p}{2p}\left(-W_{\widehat{A_\mu(p)}}^* 
W_T(0)p^\mu  p^\nu W_{\widehat{A_\nu(p)}} 
+
W_{\widehat{\phi(p)}}^*
 {\cal R}_{\phi\phi}(p)W_{\widehat{\phi(p)}} 
\right)
\nonumber\\&&
=\int \frac{d^3p}{2p}W_T(0)\left(-W_{\widehat{A_\mu(p)}}^* 
p^\mu  p^\nu W_{\widehat{A_\nu(p)}} 
+ 
W_{\widehat{\phi(p)}}^*{\cal R}_{\phi B}^2 W_{\widehat{\phi(p)}} 
\right)
\nonumber\\&&
= \int \frac{d^3p}{2p}\frac{ W_T(0)}{2}\Big [\left( ip^\mu 
W_{\widehat{A_\mu(p)}}^*
+W_{\widehat{\phi(p)}}^*
 {\cal R}_{\phi B}
\right )
\left ( 
i p^\nu W_{\widehat{A_\nu(p)}} 
 + {\cal R}_{\phi B}
W_{\widehat{\phi(p)}}
\right)
\nonumber\\&&
+\left(- ip^\mu
W_{\widehat{A_\mu(p)}}^*
+W_{\widehat{\phi(p)}}^*
 {\cal R}_{\phi B}
\right )
\left ( 
-i p^\nu W_{\widehat{A_\nu(p)}} 
 + {\cal R}_{\phi B}
W_{\widehat{\phi(p)}}
\right)
\Big]
\label{diagr.13}
\end{eqnarray}

In deriving eq. (\ref{diagr.13}) we have used eq. (\ref{nonabel.12}).

By comparing eqs. (\ref{diagr.9}-\ref{diagr.11}) with eq. 
(\ref{diagr.13}) we can identify in two terms 
in round brackets the operation 
\begin{eqnarray}
i p^\nu W_{\widehat{A_\nu(p)}} 
 + {\cal R}_{\phi B}
W_{\widehat{\phi(p)}}
=
-\int d^4 x \exp(ipx)
\Box_x 
\left . W_{B(x)}\right|_{p^2=0}.
\label{diagr.14}
\end{eqnarray}
If the unitarity equation (\ref{diagr.12}) contains more than 
one particle in the final state $n$, then we have to consider 
the reduction formula for a further momentum $q$. 
The unphysical states are described by the massless modes 
in $A_\mu$ and $\phi$. According to eq. (\ref{nonabel.2a}) we have 
\begin{eqnarray}&& 
W_{B(x)A_\mu(y)} = W_{\bar c(x)A_\mu^*(y)} 
\label{diagr.15a} 
\\&& W_{B(x)\phi(y)} = W_{\bar c(x)\phi^*(y)} . 
\label{diagr.15b} 
\end{eqnarray} 
Then from eqs. (\ref{diagr.14}) and (\ref{nonabel.5}) 
\begin{eqnarray}
i p^\nu W_{\widehat{A_\nu(p)}A_\mu(y) } 
 + {\cal R}_{\phi B}(p)
W_{\widehat{\phi(p)}A_\mu(y)}
={\cal R}_{\bar c c }(p) W_{\widehat{\bar c (p)}A_\mu^*(y)}, 
\label{diagr.16} 
\end{eqnarray} 
from where  we eventually get  the residuum of the pole at $q^2=0$ 
\begin{eqnarray}&& 
-W_T(0) q^\sigma q^\tau \left( i p^\nu 
W_{\widehat{A_\nu(p)}\widehat{A_\tau(q)} } 
 + {\cal R}_{\phi B}(p) W_{\widehat{\phi(p)}\widehat{A_\sigma(q)}}
\right)
\nonumber\\&&
={\cal R}_{\bar c c }(p){\cal R}_{  A_\sigma^*\bar c}(q) W_{\widehat{\bar c (p)}\widehat{c(q)}} \nonumber\\&& =iq^\sigma {\cal R}_{\bar c c }(p) W_{\widehat{\bar c (p)}\widehat{c(q)}}, 
\label{diagr.17} 
\end{eqnarray} 
where the last step is due to eq. (\ref{nonabel.5}). 
\par Similarly from eq. (\ref{diagr.15b}) one gets 
\begin{eqnarray} 
i p^\nu W_{\widehat{A_\nu(p)}\phi(y) } 
 + {\cal R}_{\phi B}(p)
W_{\widehat{\phi(p)}\phi(y)}
={\cal R}_{\bar c c }(p) W_{\widehat{\bar c (p)}\phi^*(y)}, 
\label{diagr.18} 
\end{eqnarray} 
where the residuum at $q^2=0$ is 
\begin{eqnarray}&& 
{\cal R}_{\phi \phi}(q) 
\left( i p^\nu W_{\widehat{A_\nu(p)}\widehat{ \phi(q)} } 
 + {\cal R}_{\phi B}(p)
W_{\widehat{\phi(p)}\widehat{ \phi(q)}}
\right)
\nonumber\\&&
={\cal R}_{\bar c c }(p){\cal R}_{  \phi^*\bar c}(q) 
W_{\widehat{\bar c (p)}\widehat{c(q)}} 
={\cal R}_{\bar c c }(p){\cal R}_{B  \phi}(q) 
W_{\widehat{\bar c (p)}\widehat{c(q)}}, 
\label{diagr.19} 
\end{eqnarray} 
where the last step is due to eq. (\ref{nonabel.5}). 
\par The first term in eq. (\ref{diagr.13}), after the insertion 
of an extra intermediate state with momentum $q$ reads 
\begin{eqnarray}&& 
\int \frac{d^3p}{2p}\int \frac{d^3q}{2q}\frac{ W_T(0)}{2} \nonumber\\&& \Big [
- W_T(0)q^\sigma
\left( ip^\mu 
W_{\widehat{A_\mu(p)}\widehat{A_\sigma(q)}}^*
+W_{\widehat{\phi(p)}\widehat{A_\sigma(q)}}^*
 {\cal R}_{\phi B}
\right )
\nonumber\\&&
q^\tau
\left ( 
i p^\nu W_{\widehat{A_\nu(p)}\widehat{A_\tau(q)}} 
 + {\cal R}_{\phi B} W_{\widehat{\phi(p)}\widehat{A_\tau(q)}}
\right)
\nonumber\\&&
+{\cal R}_{\phi \phi}(q)
\left( ip^\mu 
W_{\widehat{A_\mu(p)}\widehat{\phi(q)}}^*
+W_{\widehat{\phi(p)}\widehat{\phi(q)}}^*
 {\cal R}_{\phi B}
\right )
\nonumber\\&&
\left ( 
i p^\nu W_{\widehat{A_\nu(p)}\widehat{\phi(q)}} 
 + {\cal R}_{\phi B}
W_{\widehat{\phi(p)}\widehat{\phi(q)}}
\right)
\Big  ]
\label{diagr.20}
\end{eqnarray}
By using eqs. (\ref{diagr.17}) and (\ref{diagr.19}) we get
\begin{eqnarray}&&
= \int \frac{d^3p}{2p}\int \frac{d^3q}{2q}\frac{ W_T(0)}{2} 
\nonumber\\&& \Big [ q^\sigma \left( ip^\mu 
W_{\widehat{A_\mu(p)}\widehat{A_\sigma(q)}}^*
+W_{\widehat{\phi(p)}\widehat{A_\sigma(q)}}^*
 {\cal R}_{\phi B}
\right )
i {\cal R}_{\bar c c }(p)
W_{\widehat{\bar c (p)}\widehat{c(q)}}
\nonumber\\&&
+
\left( ip^\mu 
W_{\widehat{A_\mu(p)}\widehat{\phi(q)}}^*
+W_{\widehat{\phi(p)}\widehat{\phi(q)}}^*
 {\cal R}_{\phi B}
\right )
{\cal R}_{\bar c c }(p){\cal R}_{B  \phi}(q)
W_{\widehat{\bar c (p)}\widehat{c(q)}}
\Big  ]
\nonumber\\&&
=\int \frac{d^3p}{2p}\int \frac{d^3q}{2q}
\frac{ W_T(0)}{2} {\cal R}_{\bar c c }(p)
W_{\widehat{\bar c (p)}\widehat{c(q)}} 
\nonumber\\&& 
\Big [ ip^\mu \left( i q^\sigma W_{\widehat{A_\mu(p)}\widehat{A_\sigma(q)}}^*
+ {\cal R}_{B  \phi}(q)W_{\widehat{A_\mu(p)}\widehat{\phi(q)}}^*
\right )
\nonumber\\&&
+ 
{\cal R}_{B  \phi}(p)
\left(i q^\sigma W_{\widehat{\phi(p)}\widehat{A_\sigma(q)}}^*
+ {\cal R}_{B  \phi}(q)W_{\widehat{\phi(p)}\widehat{\phi(q)}}^*
\right )
\Big  ]
\label{diagr.21}
\end{eqnarray}
Now we use again eqs. (\ref{diagr.17}) and (\ref{diagr.19})
on the line with momentum $q$
\begin{eqnarray}&&
=\int \frac{d^3p}{2p}\int \frac{d^3q}{2q}\frac{ W_T(0)}{2} {\cal R}_{\bar c c }(p)W_{\widehat{\bar c (p)}\widehat{c(q)}} 
\nonumber\\&& 
\Big [ \left( \frac{{\cal R}_{\bar c c }^*(q)}{W_T(0)} W_{\widehat{\bar c (q)}\widehat{c(p)}}^* \right )
+ 
{\cal R}_{B  \phi}(p)
\left(
{\cal R}_{\bar c c }^*(q)
\frac{{\cal R}_{B  \phi}(q)}{{\cal R}_{\phi  \phi}(q)} W_{\widehat{\bar c (q)}\widehat{c(p)}}^* \right ) \Big  ] 
\nonumber\\&& 
=\int \frac{d^3p}{2p}\int \frac{d^3q}{2q} {\cal R}_{\bar c c }(p){\cal R}_{\bar c c }^*(q) W_{\widehat{\bar c (q)}\widehat{c(p)}}^* W_{\widehat{\bar c (p)}\widehat{c(q)}} 
\label{diagr.22} 
\end{eqnarray} 
which cancels the contribution of the ghost with momentum
$p$ and anti-ghost with momentum $q$. 
\par
The second term in eq. (\ref{diagr.13}) gives
\begin{eqnarray}&&
=\int \frac{d^3p}{2p}\int \frac{d^3q}{2q}
{\cal R}_{\bar c c }^*(p){\cal R}_{\bar c c }(q) W_{\widehat{\bar c (q)}\widehat{c(p)}} W_{\widehat{\bar c (p)}\widehat{c(q)}}^* \label{diagr.23} \end{eqnarray} 
i.e. a quantity that cancels the contribution of the ghost with momentum $q$ and anti-ghost with momentum $p$. 


\section{Operatorial approach}
\label{sec:operatorial}

Physical unitarity can also be analyzed by using a complementary operatorial
approach \cite{Curci:1976yb,Kugo:1977zq,Becchi:1985bd}.
We start by defining the S-matrix elements from the connected
generating functional by means of the following equation
\begin{eqnarray}
S = \left . : \Sigma : W  \right |_{J=\psi^*=0} 
\label{ar.1}
\end{eqnarray}
$\psi^*$ denotes collectively the antifields coupled to the fields $\psi$
in eq.(\ref{diagr.1}), $J$ the sources coupled to $\psi$. The normal
product prescription is indicated by vertical dots.

In the bosonic massless sector it is convenient to use the (overcomplete)
set of states spanned 
by $\phi, B, A_\mu$. Then the operator $\Sigma$ takes the form 
(suppressing the color indices)
\begin{eqnarray}
\Sigma & = & \exp \Big ( \int d^4p \, \big ( 
  A^\mu ~ \G_{\mu\nu} \frac{\delta}{\delta J_\nu} 
+ A^\mu ~ \G_{\mu\phi} \frac{\delta}{\delta J_\phi}
+ A^\mu ~ \G_{\mu B} \frac{\delta}{\delta J_B}
\nonumber \\
       &   & ~~~~~~~~~~
+ B ~ \G_{B\nu} \frac{\delta}{\delta J_\nu} 
+ B ~ \G_{B\phi} \frac{\delta}{\delta J_\phi}
\nonumber \\
       &   & ~~~~~~~~~~
+ \phi ~ \G_{\phi \nu} \frac{\delta}{\delta J_\nu}
+ \phi ~ \G_{\phi \phi} \frac{\delta}{\delta J_\phi}
+ \phi ~ \G_{\phi B} \frac{\delta}{\delta J_B}
\nonumber \\
       &   & ~~~~~~~~~~
+ c~ \G_{c\bar c}\frac{\delta}{\delta J_{\bar c}}
+ \bar c~ \G_{\bar c c}\frac{\delta}{\delta J_c} 
\big ) \Big ) 
\, 
\label{ar.2}
\end{eqnarray}
where use has been made of eq.(\ref{nonabel.6}).

The relevant asymptotic BRST charge $Q$ associated with the
BRST operator in eq.(\ref{nonabel.1}) is given by
\begin{eqnarray}
&& [Q,A_\mu] = \G_{cA_\mu^*} c \, , ~~~~
[Q,\phi] = \G_{c\phi^*} c \, , \nonumber \\
&& [Q,c]_{+} = 0 \, , ~~~~ [Q, \bar c]_{+} = B \, , ~~~~
[Q,B]= 0 \, .
\label{ar.3}
\end{eqnarray}
The commutator between $Q$ and the operator $:\Sigma:$ yields
\begin{eqnarray}
[Q, :\Sigma:] & = & : \int d^4p \, 
\Big [ 
~ c ~ \Big (  ( \G_{cA_\mu^*}\G_{\mu\nu} + \G_{c\phi^*}\G_{\phi\nu} )
\frac{\delta}{\delta J_\nu} \nonumber \\
&  & ~~~~~~~~~~~
+ ( \G_{cA_\mu^*}\G_{\mu\phi} + \G_{c\phi^*}\G_{\phi\phi} )
\frac{\delta}{\delta J_\phi}
\nonumber \\
&  & ~~~~~~~~~~~ + ( \G_{cA_\mu^*} \G_{\mu B} + \G_{c\phi^*}\G_{\phi B} )
\frac{\delta}{\delta J_B} \Big ) 
\nonumber \\
&  & ~~~~~~~~~~~
+ B ~ \G_{\bar c c}\frac{\delta}{\delta J_c} 
\Big ]
\Sigma :
\label{ar.4}
\end{eqnarray}
We now use the following three relations obtained
by differentiating the STI in eq.(\ref{nonabel.2b})
w.r.t. to the ghost $c$ and $A_\nu,\phi,B$ respectively:
\begin{eqnarray}
&& \G_{c A_\mu^*} \G_{\mu\nu} + \G_{c \phi^*}\G_{\nu \phi} = 0 \, , 
\label{ar.5a} \\
&& \G_{c A_\mu^*} \G_{\mu\phi} + \G_{c\phi^*} \G_{\phi\phi} = 0 \, , 
\label{ar.5b} \\
&& \G_{c A_\mu^*} \G_{\mu B} + \G_{c\phi^*} \G_{\phi B} + \G_{c \bar c} = 0 
\, . \label{ar.5c}
\end{eqnarray}

\medskip

\medskip
Then eq.(\ref{ar.4}) simplifies to
\begin{eqnarray}
[Q, : \Sigma : ] = : \int d^4p \, 
\Big [ - c ~ \G_{c\bar c}\frac{\delta}{\delta J_B} +
B ~ \G_{\bar c c}\frac{\delta}{\delta J_c} \Big ] \Sigma :\, .
\label{ar.6}
\end{eqnarray}
The above expression equals the commutator $[ : \Sigma :, {\cal S}]$
(one again needs eqs.(\ref{ar.5a})-(\ref{ar.5c})):
\begin{eqnarray}
[Q, : \Sigma :] = [ : \Sigma :, {\cal S}] \, .
\label{ar.7}
\end{eqnarray}
From the above equation it follows that $Q$ is a conserved charge:
\begin{eqnarray}
[Q, S] = [Q, \left . :\Sigma: W \right |_{J=\psi^*=0}] = 0 \, .
\label{ar.8}
\end{eqnarray}
Moreover from eq.(\ref{ar.3}) $Q$ is nilpotent. 

\medskip
Now we are in a position to characterize the Hilbert space
${\cal H}_{\rm phys} = {\rm Ker } ~ Q / {\rm Im } ~ Q$.
From eq.(\ref{ar.3}) the three massive states of the gauge field
$A_\mu$ belong to ${\cal H}_{\rm phys}$.
The massless unphysical states can be analyzed as follows.
$B$ is in the kernel of $Q$ but is $Q$-exact, as it follows
from eq.(\ref{ar.3}). Moreover the state $X$ in eq.(\ref{nonabel.13})
does not belong to the kernel of $Q$. Finally 
$Y_\mu$ in eq.(\ref{nonabel.17}) is in the kernel of $Q$,
since
\begin{eqnarray}
[Q, Y_\mu]  & = & a \Big ( \G_{A^*_\mu c} - \frac{1}{p^2} 
\frac{1}{W_{\phi B}} i p_\mu \G_{\phi^* c} \Big ) c 
\nonumber \\
& = & \frac{a}{W_{\phi B}} \Big ( W_{\phi B}\G_{A^*_\mu c}
- \frac{i p_\mu}{p^2} \G_{\phi^* c} \Big ) c
\nonumber \\
& = & 
 \frac{a}{p^2 W_{\phi B}} \frac{ (-i p_\mu)}{\G_{\phi\phi}}
\Big ( \G_{A_\nu \phi(-p)} \G_{A^*_\nu (-p) c(p)} + \G_{\phi \phi(-p)}
\G_{\phi^*(-p)c(p)} \Big ) c 
\nonumber \\
& = &  0 
\label{ar.9}
\end{eqnarray}
by virtue of eq.(\ref{ar.5b}). However such a linear combination
does not have a pole at $p^2=0$, as it follows from the analysis
of Sect.~\ref{sec:nonabel}.
$\bar c$ is not in the kernel of $Q$, while $c$ forms a BRST doublet
together with $\phi$ (see eq.(\ref{ar.3})).

We conclude that the only physical states are given by the
three transverse massive polarizations of the gauge fields $A_\mu$.

\section{Conclusions}
\label{sec:conclusions}
Under the assumption that the theory 
can be defined in some subtraction scheme
fulfilling the ST identities,
the ghost equation and the B-equation, 
in the Landau gauge
the unphysical pole
of the vector meson propagator is a single pole located at $p^2=0$.
Being a consequence of the symmetries of the theory, this result
holds irrespective of the intricacies of the subtraction operation
to be envisaged in the context of a non power-counting renormalizable
theory.
We notice that if additional singularities 
(beyond the single pole of the physical
massive states) are generated in $W_T$  by the subtraction scheme,
they influence the (asymptotic) 2-point correlation function of $F_{\mu\nu}^a$ 
and therefore affect the  physical observables of the theory.

\medskip

\medskip
A detailed study of physical unitarity has
been carried out within this framework. 
We find that in the Landau gauge physical unitarity is fulfilled
provided that the normalization condition in eq.(\ref{int.3}) 
is imposed. This is 
an all-order universal constraint on candidate subtraction schemes 
(like dimensional regularization \cite{Appelquist:1980vg}) required
in order 
to achieve a consistent quantum definition of the St\"uckelberg model.

\section*{Acknowledgments}

One of us (RF) is honored to  gratefully
acknowledge the warm hospitality and the partial financial support 
of the Max-Planck-Institut f\"ur Physik Werner-Heisenberg-Institut, 
where part of this work has been accomplished.
Valuable discussions with and useful comments from D.~Maison are also
gratefully acknowledged.

\appendix

\section{Other covariant gauges}
\label{app:other}

We briefly consider also the case of general covariant gauges
given by the gauge fixing term
\begin{eqnarray}
S_{g.f.} = \int d^4 x ~2\,Tr\left( \frac{\alpha}{2}B^2 +
B\partial^\mu A_\mu - \bar c \partial^\mu D[A]_\mu c
\right).
\label{other.1}
\end{eqnarray}
The Slavnov-Taylor identities (\ref{nonabel.2a}) and (\ref{nonabel.2b})
as well as the ghost equations (\ref{nonabel.4a}) and
(\ref{nonabel.4a}) unchanged. The $B$-field equation is now
\begin{eqnarray}&&
\alpha W_B +J_{B}+\partial^\mu W_{A^{\mu}}=0
\label{other.3a}
\\&&
\Gamma_{B}=\alpha B+ \partial^\mu A_{\mu}.
\label{other.3b}
\end{eqnarray}

Consequently eqs. (\ref{nonabel.5}) become
\begin{eqnarray}
&& W_{A^\mu B} = -i\frac{p_\mu}{p^2}, \quad
W_{A^\mu \phi} =-i\alpha ~\frac{p_\mu}{p^2}W_{B\phi}, \quad
W_{B B} = 0, \quad
\nonumber \\&& 
W_{B \phi} = W_{\phi^* \bar c}, \quad 
W_{A^{*\mu}\bar c}= i\frac{p_\mu}{p^2} 
\label{other.5} 
\end{eqnarray} 
and similarly 
\begin{eqnarray}&& 
\Gamma_{B A^\mu } = -i p_\mu, \quad 
\Gamma_{B \phi} =0 , \quad 
\Gamma_{B B} = \alpha  , \quad 
\nonumber \\&& 
\Gamma_{\bar c c} =  -i p^\mu \Gamma_{A^{*\mu}c} , \quad 
\Gamma_{A_\mu\phi}\Gamma_{A^{*\mu}c} + \Gamma_{\phi\phi}\Gamma_{\phi^*c} = 0.
\label{other.6}
\end{eqnarray}
We impose the relation between the two-point functions 
$W$ and $\Gamma$ as in eq. (\ref{nonabel.8}) and thus get
some conditions  as in eqs. (\ref{nonabel.9a}
-\ref{nonabel.10}).
\begin{eqnarray}&& 
W_{A^\mu A^\rho}\Gamma_{A_\rho A^\nu } 
+W_{A^\mu B}\Gamma_{B A^\nu }+W_{A^\mu \phi}\Gamma_{\phi A^\nu }
= -g_{\mu\nu}
\nonumber\\&&
\Longrightarrow W_T \Gamma_T = -1,  \quad W_L \Gamma_L
-i\alpha W_{B\phi}\frac{1}{p^2}p^\nu\Gamma_{\phi A^\nu }=0
\label{other.8a} \\&& 
W_{A^\mu A^\rho}\Gamma_{A_\rho B} 
+W_{A^\mu B}\Gamma_{B B }=0 
\nonumber\\&&
\Longrightarrow   W_L 
-\alpha \frac{1}{p^2} =0
\label{other.8b} \\&& 
W_{A^\mu A^\rho}\Gamma_{A_\rho \phi} 
+W_{A^\mu \phi}\Gamma_{\phi \phi }=0 
\nonumber\\&&
\Longrightarrow W_L p^\rho\Gamma_{A^\rho \phi } -i\alpha
W_{B\phi}\Gamma_{\phi\phi}=0
\label{other.8c} \\&& 
W_{B A^\rho}\Gamma_{A_\rho A^\mu } +W_{ B\phi}\Gamma_{\phi A^\mu }=0 
\nonumber\\&&
\Longrightarrow
 i \Gamma_L 
+  W_{ B\phi}p_\mu\Gamma_{\phi A^\mu }=0 
\label{other.8d} \\&& 
W_{B A^\rho}\Gamma_{A_\rho B } +W_{ B\phi}\Gamma_{\phi B }= -1 
\label{other.8e} \\&& 
W_{ BA_\mu}\Gamma_{A^\mu\phi} + W_{B \phi}\Gamma_{\phi \phi}=0 
\nonumber\\&&
\Longrightarrow i\frac{p_\mu}{p^2}\Gamma_{A^\mu\phi}+
W_{B \phi}\Gamma_{\phi \phi}=0 
\label{other.8f} \\&& 
W_{\phi A_\nu}\Gamma_{A^\nu A^\mu}+
W_{\phi B}\Gamma_{BA^\mu} + W_{\phi \phi}\Gamma_{\phi A^\mu}=0 
\nonumber\\&& 
\Longrightarrow 
i\alpha  W_{B\phi}\Gamma_L
-i W_{ B\phi} p^2 + W_{\phi \phi}p_\mu\Gamma_{\phi A^\mu}=0 
\label{other.8g} \\&& 
W_{\phi A^\rho}\Gamma_{A_\rho B } +W_{\phi B}\Gamma_{B B}
+W_{\phi \phi}\Gamma_{\phi B}=0
\label{other.8h}
\\&&
W_{\phi A^\rho}\Gamma_{A_\rho \phi } +W_{\phi B}\Gamma_{B \phi}
+W_{\phi \phi}\Gamma_{\phi \phi}=-1 
\nonumber \\&& 
\Longrightarrow
 i\alpha \frac{p^\rho}{p^2} W_{B \phi}\Gamma_{A_\rho \phi } 
+ W_{\phi \phi}\Gamma_{\phi \phi}=-1.
\label{other.8i}
\end{eqnarray}
From the eqs. (\ref{other.8a}-\ref{other.8i}) we get
\begin{eqnarray}&&
W_T = - \frac{1}{\Gamma_T}
\label{other.9}\\&&
W_L = \frac{\alpha}{p^2}
\label{other.9a}\\&&
\Gamma_{L} = p^2 \Gamma_{\phi \phi} W_{B \phi}^2
\label{other.9b}\\&&
\Gamma_{ \phi A_\rho} = -i  p_\rho\Gamma_{\phi \phi} W_{B \phi}
\label{other.9c} \\&&
W_{\phi \phi}\Gamma_{\phi \phi}=-1 + \frac{\alpha}{p^2}\Gamma_{L}.
\label{other.9d} 
\end{eqnarray} 
Eq. (\ref{other.9a}) shows that the connected two-point function
$W_{A_\mu A_\nu}$   
potentially
develops a double pole at $p^2=0$.

\medskip
From eq.(\ref{other.9d}), (\ref{other.9b}) and eq.(\ref{other.9c}) we get
\begin{eqnarray}
W_{\phi\phi} & = & \frac{1}{\G_{\phi\phi}} \Big (
-1  + \frac{\alpha}{p^2} \G_L \Big ) \nonumber \\
& = & \frac{1}{\G_{\phi\phi}} \Big ( -1 + \alpha \G_{\phi\phi} W_{B\phi}^2
\Big ) \nonumber \\
& = &  \frac{1}{\G_{\phi\phi}} \Big ( 
-1 - \frac{\alpha}{p^2\G_{\phi\phi} }  (p^\rho \G_{\phi A_\rho})^2 \Big ) \, .
\label{other.10}
\end{eqnarray}
We define
\begin{eqnarray}
\G_{\phi A_\mu} = i p_\mu f(p^2) \, , ~~~~
\G_{A_\mu^*\phi} = i p_\mu g(p^2) \, .
\label{other.11}
\end{eqnarray}
Eq.(\ref{other.10}) becomes
\begin{eqnarray}
W_{\phi\phi} & = & \frac{1}{\G_{\phi\phi}} \Big (
-1 + \frac{\alpha}{\G_{\phi\phi}} f^2(p^2) \Big ) \nonumber \\
& = &  \frac{\alpha f^2(p^2) - \G_{\phi\phi} }{(\G_{\phi\phi})^2}
\, .
\label{other.12}
\end{eqnarray}
Then from eq.(\ref{other.6}) 
\begin{eqnarray}
\G_{\phi\phi} = - \frac{\G_{\phi A_\mu} \G_{A_\mu^*c}}{\G_{\phi^*c}} =
-\frac{p^2 f(p^2) g(p^2)}{\G_{\phi^*c}} \, .
\label{other.14}
\end{eqnarray}
By assuming that $\G_{\phi^* c}$ tends to a constant different from zero
for $p^2 \rightarrow 0$ $\G_{\phi\phi}$ has a zero at $p^2=0$. 
At tree-level it is the only zero of $\G_{\phi\phi}$.
Provided that this is true also at the quantum level, 
we see from eq.(\ref{other.12}) that   $W_{\phi\phi}$ also 
potentially
(i.e. by excluding that $f(p^2)$ has a zero in $p^2=0$)
develops a double pole at $p^2=0$.


%

%

%
\bibliography{reference}

\begin{thebibliography}{99}
\small

\bibitem{ym}
C.~N.~Yang and R.~L.~Mills,
Phys.\ Rev.\  {\bf 96} (1954) 191.

\bibitem{Higgs:1964ia}
P.~W.~Higgs,
Phys.\ Lett.\  {\bf 12} (1964) 132, Phys.\ Lett.\  {\bf 13} (1964)
508, Phys.\ Rev.\  {\bf 145} (1966) 1156. \\
F.~Englert and R.~Brout,
Phys.\ Rev.\ Lett.\  {\bf 13} (1964) 321.
G.~S.~Guralnik, C.~R.~Hagen and T.~W.~B.~Kibble,
Phys.\ Rev.\ Lett.\  {\bf 13} (1964) 585.
T.~W.~B.~Kibble,
Phys.\ Rev.\  {\bf 155} (1967) 1554.


\bibitem{stueckelberg} E.~C.~G.~ St\"uckelberg, Helv.\ Phys.\ Helv.\ Acta
{\bf 11} (1938), 299.

\bibitem{ruegg} For a recent review see 
H.~Ruegg and M.~Ruiz-Altaba, ``The Stueckelberg field,''
arXiv:hep-th/0304245.

\bibitem{Esole:2004rx}
M.~Esole,
``The non-local massive Yang-Mills action as a gauged sigma model,''
arXiv:hep-th/0407069.

\bibitem{Slavnov:1970tk}
A.~A.~Slavnov and L.~D.~Faddeev,
Theor.\ Math.\ Phys.\  {\bf 3} (1970) 312
[Teor.\ Mat.\ Fiz.\  {\bf 3} (1970) 18].

\bibitem{Shizuya:1975ek}
K.~Shizuya,
Nucl.\ Phys.\ B {\bf 94} (1975) 260.

\bibitem{Grosse-Knetter:1993nu}
C.~Grosse-Knetter,
Phys.\ Rev.\ D {\bf 48} (1993) 2854
[arXiv:hep-ph/9304310].

\bibitem{Banerjee:1997sf}
R.~Banerjee and J.~Barcelos-Neto,
Nucl.\ Phys.\ B {\bf 499} (1997) 453
[arXiv:hep-th/9701080].

\bibitem{Dragon:1996tk}
N.~Dragon, T.~Hurth and P.~van Nieuwenhuizen,
Nucl.\ Phys.\ Proc.\ Suppl.\  {\bf 56B}, 318 (1997)
[arXiv:hep-th/9703017].

\bibitem{delbourgo} 
see R.~Delbourgo, S.~Twisk and G.~Thompson,
Int.\ J.\ Mod.\ Phys.\ A {\bf 3} (1988) 435 
and references therein.

\bibitem{Becchi:1975nq}
C.~Becchi, A.~Rouet and R.~Stora,
Annals Phys.\  {\bf 98}, 287 (1976).

\bibitem{Becchi:1974xu}
C.~Becchi, A.~Rouet and R.~Stora,
Phys.\ Lett.\ B {\bf 52} (1974) 344.

\bibitem{Curci:1976yb}
G.~Curci and R.~Ferrari,
Nuovo Cim.\ A {\bf 35}, 273 (1976).


\bibitem{Kugo:1977zq}
T.~Kugo and I.~Ojima,
Phys.\ Lett.\ B {\bf 73}, 459 (1978),
Progr.\ Theor.\ Phys. \ {\bf 60}, 1869 (1978).

\bibitem{Becchi:1985bd}
C.~Becchi,
``Lectures On The Renormalization Of Gauge Theories,''
in *Les Houches 1983, Proceedings, Relativity, Groups and Topology, II*, 
787-821.



\bibitem{Nakanishi:sm}
N.~Nakanishi,
Phys.\ Rev.\ D {\bf 5} (1972) 1324,
Prog.\ Theor.\ Phys.\ Suppl.\  {\bf 35} (1966) 1111,
Prog.\ Theor.\ Phys.\ Suppl.\  {\bf 51} (1972) 1.

\bibitem{lautrup} 
B. Lautrup, Mat. Fys. Meeld. Dan. Viol. Selsk. 35 (1967) no. 11.


\bibitem{st} 
 J.~C.~Taylor, Nucl.\ Phys.\ {\bf B33} (1971) 436; \\
 A.A.Slavnov, {Theor.\ Math.\ Phys.\ }  {\bf 10} (1972) 99.

\bibitem{fp} L.~D.~ Faddeev and V.~N.~ Popov, Phys.\ Lett.\ B {\bf 25} 
(1967) 29.

\bibitem{Boulware} D.~ G.~ Boulware, Ann.\ Phys.\ (N.Y.)\ {\bf 56}
  (1970) 140.


\bibitem{'tHooft:1971rn}
G.~'t Hooft,
Nucl.\ Phys.\ B {\bf 35}, 167 (1971).
\\
G.~'t Hooft and M.~J.~G.~Veltman,
Nucl.\ Phys.\ B {\bf 44}, 189 (1972), 
Nucl.\ Phys.\ B {\bf 50}, 318 (1972).


\bibitem{BRST}
See Ref. \cite{Becchi:1974xu} 
and   L.V. Tyutin, Lebedev preprint FIAN n.39 (1975).


\bibitem{Cutkosky:1960sp}
R.~E.~Cutkosky,
J.\ Math.\ Phys.\  {\bf 1} (1960) 429.

\bibitem{Veltman:1963th}
M.~J.~G.~Veltman,
Physica {\bf 29}, 186 (1963).

\bibitem{Appelquist:1980vg}
T.~Appelquist and C.~W.~Bernard,
Phys.\ Rev.\ D {\bf 22} (1980) 200.

\bibitem{zj}
J.~Zinn-Justin, {\em Renormalization of gauge theories}, lectures given
at {\em International Summer Institute for Theoretical Physics}, Bonn,
Germany, Jul. 29 - Aug. 9, 1974, published in Bonn Conf. 1974,2.

\bibitem{Gomis:1994he}
J.~Gomis, J.~Paris and S.~Samuel,
Phys.\ Rept.\  {\bf 259} (1995) 1
[arXiv:hep-th/9412228].

%
%

\bibitem{Henneaux:1998hq}
M.~Henneaux and A.~Wilch,
Phys.\ Rev.\ D {\bf 58}, 025017 (1998)
[arXiv:hep-th/9802118].

\normalsize
\end{thebibliography}

\end{document}